\documentclass[aps,prl,amssymb,amsfonts,twocolumn]{revtex4}

\usepackage{graphicx}

\begin{document}

\title{Landau--Zener interferometry for qubits}

\author{A.~V.~Shytov$^{1,2}$, D.~A.~Ivanov$^{3,}$\footnote{Present 
address: Paul Scherrer Institut, CH-5232 Villigen PSI, Switzerland} and
M.~V.~Feigel'man$^1$ }

\address{
$^1$L.\ D.\ Landau Institute for Theoretical Physics, Moscow
117940, Russia\\
$^2$Institute for Theoretical Physics, University of California at Santa
Barbara, USA \\
$^3$Institute for Theoretical Physics, ETH-Z\"urich,  CH-8093, Switzerland
}

\begin{abstract}
One may probe coherence of a qubit by periodically sweeping its control
parameter. The qubit is then excited by the Landau--Zener (LZ) mechanism.
The interference between multiple LZ transitions leads to an 
oscillatory dependence of the energy absorption rate on the sweeping 
amplitude and on the period. This interference pattern allows to determine
the decoherence time of the qubit. We introduce 
a simple phenomenological model describing this ``interferometer'', 
and find the form of the interference pattern.  
\end{abstract}

\maketitle

During the last few years, a number of proposals for constructing 
quantum bits (qubits) from  mesoscopic Josephson junctions have
appeared~\cite{Schoen,Mooij1,Ioffe,pijunc} and first experimental 
results in this direction have been reported~
\cite{nakamura,mooij,nakamura-mooij,luckens,vion,martinis,han}.
Large part of these qubits are actually different physical realizations of an
externally controllable quantum double-well system, with nearly equal depths
$E_{1,2}$ of both wells $|E_1-E_2| \ll \omega_0 \ll E_1$
(here $\omega_0$ is the oscillation frequency within a single well),
and with the inter-well tunneling  amplitude 
$\Delta \sim |E_1-E_2|$.
The above conditions ensure that higher eigenstates of the system are
separated from the nearly degenerate doublet by a large gap  
(compared to $\Delta$), and
the probability of their excitation can be neglected.
The energy difference
$|E_1-E_2|$ is controlled by an external time-dependent  parameter $x(t)$,
 which is either voltage
for the SET-based ``charge'' qubit~\cite{Schoen,nakamura}, or magnetic flux 
through the Josephson junction loop for the ``phase''
qubit~\cite{Ioffe,pijunc,mooij,luckens}.
A review of recent results for both types of superconductive qubits
can be found in~\cite{Makhlin1}.
Quantum  manipulations with qubits
involve varying in time both $x(t)$ and  $\Delta(t)$ (as well as more
complicated two-qubit manipulations).
Since the overall scale of possible $E_{1,2}$ variations as function of
control parameter $X$ is very large compared to
relevant values of  $\Delta$ (e.g. it was  more than 100 times larger in the
design of Ref.~\cite{Mooij1}),  
fluctuations of $x(t)$  are expected to be one of the most
important sources of dephasing in such qubits~\cite{pijunc}. 
Indeed, the experimental data of Ref.~\cite{mooij}
seem to confirm these expectations.
 
Thus the first problem to be addressed in the development of 
this type of qubits is
to find  a convenient probe which tests whether the device
%produces coherent Rabi oscillations. 
undergoes coherent evolution. 
Resonant absorption method was used in
experiments of Refs.~\cite{mooij,luckens}, 
whereas Nakamura {\it et al} have used
time-domain manipulations~\cite{nakamura}. 
%In some cases 
It may be preferable to employ
simpler methods to measure decoherence time of a qubit, without
super-high-frequency (in the GHz range) manipulations. 
%at the first stage of the device characterization.  
One such method (based on the measurement of
static nonlinear Andreev conductance) was proposed
in Ref.~\cite{andreevfork} 
for the specific case of superconducting phase qubit like those proposed
in Refs.~\cite{Ioffe,pijunc}. Another low-frequency probe is the observation
of Ramsey fringes, which was performed for superconducting qubits 
in~\cite{nakamura-mooij,vion}.
%In~\cite{vion,nakamura-mooij}, 
%Ramsey fringes were observed in superconducting qubits. 
In the present paper we propose and analyze a different method
to determine the decoherence  time of a qubit by a low-frequency
non-resonant measurement.   

\begin{figure}
\includegraphics[scale=0.4]{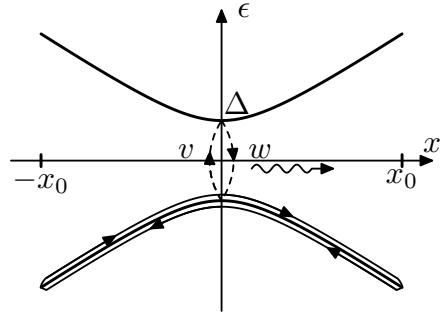}
\caption[]{
Eigenvalues $\epsilon$ of the Hamiltonian (\ref{hamiltonian})
depend on the control parameter $x$.
When the control parameter passes the Landau--Zener point ($x{=}0$), 
the qubit may be non-adiabatically excited with the probability $v$. 
The upper state may also decay into the lower one, dissipating the
energy into the environment. The decay probability is $w$.
}
\label{fig1}
\end{figure}

The idea is to employ the Landau--Zener (LZ) non-adiabatic
tunneling processes. If  the control parameter $x(t)$ 
changes in such a way that
the ``collision region'' with $\Delta \sim |E_1-E_2|$ is traversed, 
the system 
may become excited from the lower to the upper level. 
If the parameter $x(t)$ is changed periodically, successive Landau--Zener
tunneling events interfere, and such an interference
allows to estimate the decoherence in the qubit.
The interference picture is easiest to describe in the setup 
with the amplitude of the parameter
sweep $x_0$ large compared to the width of the ``transition region''
(Fig.~\ref{fig1}). If the dephasing time $t_{\rm deph}$ is larger
than the period of the parameter sweep $t_0$, the interference
between successive Landau--Zener transitions has an
oscillatory dependence upon $x_0$ and $t_0$. The interference may
be observed from the rate of energy dissipation which is proportional
to the occupation of the upper level. The precise form and the
strength of the interference pattern is determined by the interplay
of the coherent Landau--Zener transitions and of the spontaneous
incoherent decay $E_2 \to E_1$. 
Performing interference experiments at a constant offset $x_{\rm off}$
in the oscillations of the parameter $x(t)$ and at different values
of $x_0$ and $t_0$ allows to determine the rates of dephasing and
of inelastic relaxation in the qubit.

Specifically, we model the qubit by the two-level Hamiltonian
\begin{equation}
H= 
\left(
    \begin{array}{cc}
       x(t) + \hat{\xi}(t)  & \Delta          \\
       \Delta         & - x(t) - \hat{\xi}(t)
    \end{array}
\right).
\label{hamiltonian}
\end{equation}
In this paper, we consider only one mode of qubit operation, namely
varying the diagonal controlling parameter $x(t)$ while keeping the
gap $\Delta$ fixed. We also assume that the main source of
decoherence are thermal and quantum fluctuations 
of the control parameter around its intended
value. Such fluctuations are described by the operator $\hat{\xi}(t)$.
Phase fluctuations of $\Delta$ may also be incorporated into $\hat{\xi}(t)$
by an appropriate gauge transformation.

An example of the physical system leading to the Hamiltonian
(\ref{hamiltonian}) is the superconducting phase qubit 
\cite{Mooij1,mooij,andreevfork}.
The control parameter $x(t)$ %in (\ref{hamiltonian})
in this design is the magnetic flux through
the qubit loop. 
The diagonal coupling to the environment 
is realized via coupling to the magnetic flux 
(including the fluctuations of the external
electromagnetic field~\cite{pijunc,Levitov-decoherence} and coupling to
nuclear magnetic spins \cite{nuclear-spins}). We also remark here 
that the diagonal coupling to the environment in the 
Hamiltonian~(\ref{hamiltonian}) does not represent the most 
general form of coupling. In general, off-diagonal coupling may also be
present, which would lead to the inelastic decay of the qubit states 
away from the Landau--Zener transition region. However, we assume that the
off-diagonal coupling is already sufficiently suppressed: 
the qubit can preserve its occupation-number information in the 
``idle'' regime (away from the Landau--Zener transition region), 
and the decoherence is determined by the diagonal coupling channel. 

The quantum variable $\hat{\xi}(t)$ 
describes coupling to the collective degree of freedom 
of the external reservoir.
It is a quantum variable corresponding to a collective degree
of freedom of the reservoir. The usual model for the reservoir
is an ensemble of harmonic oscillators\cite{Caldeira-Leggett}. Our treatment
will be phenomenological and not involving the microscopic properties
of the reservoir, therefore we do not explicitly include the reservoir
in the Hamiltonian~(\ref{hamiltonian}). However, we check the validity
of our approach by comparing it to the microscopic calculation for 
the oscillator bath model (see Appendix). 

Depending on the experimental conditions, the temperature of the
reservoir may be either lower or higher than the gap $\Delta$.
We first consider the case of the reservoir temperature much smaller
than the gap $\Delta$, and later explain how the results are modified
at higher temperature. Independently of the relation to the gap $\Delta$,
we assume that the reservoir temperature is always higher than the
sweep frequency $t_0^{-1}$: this is necessary for our treatment of
dephasing as a Gaussian noise and for our assumption of independent
dephasing processes on different half-periods of the parameter sweep.

The effect of the coupling to $\hat{\xi}(t)$ is twofold. 
In the transition region ($x(t)\sim\Delta$),
this coupling has non-vanishing matrix elements between the two
adiabatic levels, and therefore leads to inelastic transitions
between the levels. In the limit of the reservoir temperature
much lower than $\Delta$, the transitions occur mostly from the
upper to the lower level, thus attenuating the transition 
probability~\cite{Gefen1}.
Away from the transition region ($|x(t)| \gg \Delta$), 
the Hamiltonian (\ref{hamiltonian}) is almost diagonal, 
and the effect of $\hat{\xi}(t)$ is dephasing. 

The control parameter $x(t)$ is swept periodically, with
the amplitude $x_0$ and with the period $t_0$:
\begin{equation}
x(t) = x_0 \sin \frac{2\pi t}{t_0} + x_{\rm off} \ . 
\label{periodic-x}
\end{equation}
For simplicity, we first consider in detail the case of zero offset
$x_{\rm off}=0$, and then discuss the interference pattern at
arbitrary $x_{\rm off}$.

Each time the control parameter passes the Landau--Zener point
$x(t) = 0$, the Landau--Zener tunneling occurs. This tunneling is
a quantum-mechanical process  sensitive to the relative phase
of the two states. Therefore, the energy absorption per period
depends on the phase $\varphi_n$ picked up far from LZ point. 
The latter phase is determined by both the sweep amplitude and the frequency
(assuming $x_{\rm off}=0$):
\begin{equation}
\varphi_n = 2 \int\limits_{0}^{t_0/2} 
\left(x(t)+\xi(t)\right)\, dt = \varphi+\delta\varphi_n\, ,
\quad
\varphi=\frac{2 x_0 t_0}{\pi}. 
\end{equation}
Here $\varphi$ is the average phase picked up per half-period, and
$\delta\varphi_n$ are its fluctuations. If we assume that the
correlation time of $\hat{\xi}(t)$ is much shorter than $t_0$
(which is equivalent to assuming that the reservoir temperature
is much higher than $t_0^{-1}$), 
the probability distributions of $\delta\varphi_n$ are
Gaussian  and uncorrelated on different half-periods 
(labeled by the integer $n$).

The amplitude $x_0$ is assumed to be much larger then the level-crossing
region ($x_0\gg\Delta$), and the period $t_0$ should be sufficiently large,
so that the Landau--Zener transition probability \cite{Landau-Zener}
is small:
\begin{equation}
v=\exp\left(- {\Delta^2 t_0 \over 2 x_0}\right) \ll 1 \ .
\label{LZ-transition-probability}
\end{equation}
Also, $x_0$ should not be too large so that the Hamiltonian
(\ref{hamiltonian}) would still adequately describe the system. 
(For the superconducting phase qubit it implies that the 
amplitude of the flux modulation should be small compared to superconducting 
flux quantum.)

We describe the system evolution in terms of the two-level
density matrix $\rho$. The evolution per one half-period of the
parameter variation (\ref{periodic-x}) is given by the master
equation which includes the three
effects: the coherent Landau--Zener transitions, the
inelastic decay of the qubit, and the phase picked up during
the system evolution away from the level-crossing region.

We separate the decoherence effect
into the two parts: the dephasing away from the transition region
and the inelastic relaxation in the transition region. 
This separation is possible if the transition region is narrow: 
$\Delta\ll x_0$. 
Instead of microscopically deriving the relevant couplings 
(see e.~g.\ Ref.~\cite{Gefen1,Gefen2}), 
we include them phenomenologically
as independent parameters in the master equation on the 
two-level density matrix. Both types of decoherence are assumed to
be small. More precisely, we describe the decoherence by the
two dimensionless parameters: the average phase fluctuation per one sweep,
 $u = \langle \delta\varphi^2_n \rangle$, 
and the probability of inelastic decay
per one crossing of the transition region $w$.
Our parameters $w$ and $u$ are proportional to
the longitudinal and transverse relaxation rates,
respectively (defined as $\Gamma_{\rm relax}$ and $\Gamma_{\phi}$ 
in Refs.\ \cite{Makhlin1,Makhlin2}). These relaxation
rates have contributions from different decoherence channels,
their calculation from microscopic principles is difficult,
and {\em a priori} there is no universal relation between them.
Therefore instead of computing them microscopically, we introduce
them phenomenologically as two independent parameters $u,w\ll 1$.

The three small parameters $u$, $w$, and $v$ depend, in principle, on the
period and amplitude of the oscillations of $x(t)$. We shall return to 
this dependence in the end of the paper. 

The Landau--Zener transition, in the absence of inelastic events,
is described by the unitary rotation of the density matrix:
\begin{equation}
\rho \mapsto S \rho S^{\dagger}\, , 
\qquad
S = 
\left(
\begin{array}{cc}
    r     & t \\
  - t^{*} & r^{*}
\end{array}
\right) \ ,
\label{elastic}
\end{equation}
where $t$ and $r$ are the amplitudes of the transition and of
staying at the same adiabatic level, respectively. 
Amplitudes $r$ and $t$ satisfy unitarity condition 
$|r|^2+|t|^2=1$. The magnitude of $t$ determines the LZ transition
probability (\ref{LZ-transition-probability}): $|t|^2=v$. 

Let us assume first that the reservoir temperature is much
lower than the gap $\Delta$. Then
the inelastic relaxation processes may be described
as attenuation of the amplitude of being in
the upper energy state (the case of higher reservoir
temperatures will be discussed later, see Eqs.\ (\ref{inelastic-3})
and (\ref{inelastic-4}) below):
\begin{equation}
\rho \mapsto V_w\rho V_w + w \rho^{(0)} \rho_{22} \ , 
\label{inelastic}
\end{equation}
where
\begin{equation}
V_w = \left(
\begin{array}{cc}
    1    & 0 \\
    0    & \sqrt{1-w}
\end{array}
\right)\ , 
\qquad
\rho^{(0)} = 
\left( 
\begin{array}{cc}
1 & 0  \\
0 & 0
\end{array}
\right) .
\label{inelastic-2}
\end{equation}
The equilibrium density matrix $\rho^{(0)}$ describes 
the system in the lower energy state. The quantity $\rho_{22}$
is the diagonal element of the density matrix corresponding to
the upper state. 
The first term in Eq.~(\ref{inelastic}) describes the decay of the 
upper state. It attenuates the amplitudes of being in the upper energy 
state by $\sqrt{1-w}$. Thus, the probability that the upper state 
will not decay is $1-w$, and this allows to identify $w$ as the 
inelastic decay  probability. 
The second term in Eq.~(\ref{inelastic}) describes the probability flow
into the lower state due to the inelastic decay. This term is diagonal
since the decay is assumed to be incoherent. 
Note that the Eq.~(\ref{inelastic}) preserves the trace of 
the density matrix. 

A complete solution of the dissipative dynamics requires
simultaneously taking into account the Landau--Zener processes
(\ref{elastic}) and the dissipative processes (\ref{inelastic})
similarly to the treatment in Refs.~\cite{Gefen1,Gefen2},
which we do not attempt here.
Instead, note that if the transition rates are small ($v, w \ll 1$), 
the elastic and inelastic processes may be considered independently. 
Thus, we simply combine Eqs. (\ref{elastic}) and (\ref{inelastic}):
\begin{equation}
\label{combined}
\rho \mapsto S V_w \rho V_w S^{\dagger} + w \rho_{22} \rho^{(0)}\ .
\end{equation}
The matrices $V_w$ and $S$ do not commute, but the leading terms
in $w$ and $v$ do not depend on the order of multiplication. 
We ordered $S$ and $V_w$ in Eq. (\ref{combined})
so that the trace of the density matrix is conserved.

Finally, the phase picked up far from the Landau--Zener point
produces the relative phase rotation of the upper and lower
states:
\begin{equation}
\rho \mapsto \Phi_n \rho \Phi_n^{\dagger}\ , \qquad
\Phi_n = \exp\left(  i \varphi_n \sigma_{z} / 2 \right),
\label{phase}
\end{equation}
where $\sigma_z$ is the Pauli matrix. 

In this way, the evolution of the density matrix per one
sweep is described by the master equation
\begin{equation}\rho \mapsto 
\Phi_n
\left(S V_w \rho V_w S^{\dagger} + w \rho_{22} \rho^{(0)}\right) 
\Phi_n^{\dagger}
\label{master-equation}
\end{equation}

We parameterize the density matrix as 
\begin{equation}
\rho = \frac{1}{2}  \left(a_0 + {\bf a} {\bf \sigma}\right)\ . 
\end{equation}
The scalar part $a_0=1$ remains constant, as required by
the normalization of the density matrix. Then the dynamics is
described by an equation for the polarization vector ${\bf a}$.
The phase factor $\Phi_n$ in (\ref{phase})
rotates the vector ${\bf a}$ by the angle $\varphi$  about $z$-axis.
Similarly, the scattering matrix $S$  in (\ref{elastic})
rotates the vector ${\bf a}$ around some axis in $xy$-plane.
One may redefine the phases of the upper and lower states 
so that $S$ describes the rotation about $x$ axis. 
(This transformation shifts all phases $\varphi_n$ by a constant.)
Thus the polarization vector ${\bf a}$ evolves per one sweep as
\begin{equation}
{\bf a}_{n+1}= Q_n {\bf a}_{n} + w \hat{\bf z}\ ,
\label{evolution}
\end{equation}
where $Q_n$ is
\begin{equation}
Q_n=R_z(\delta\varphi_n) R_z(\varphi) R_x(\theta) A_w\ .
\label{U-n}
\end{equation}
Here $R_z$ and $R_x$ are the rotation operators about $z$- and
$x$-axes, and $A_w$ describes the attenuation:
\begin{eqnarray}
R_{z}(\varphi) &=& 
\left(
\begin{array}{ccc}
  \cos\varphi &  -\sin\varphi  & 0  \\
  \sin\varphi &   \cos\varphi  & 0  \\
  0           &   0           & 1
\end{array}
\right)\ , 
\nonumber\\
R_{x}(\theta) &=& 
\left(
\begin{array}{ccc}
  1 & 0          &           0 \\
  0 & \cos\theta & -\sin\theta \\
  0 & \sin\theta &  \cos\theta
\end{array}
\right)\ ,
\label{R-A}
\\
A_w &=& \left(
\begin{array}{ccc}
\sqrt{1-w} & 0 & 0 \\
0 & \sqrt{1-w} & 0 \\
0 & 0 & 1-w 
\end{array}
\right)\ . 
\nonumber
\end{eqnarray}
The parameter $\theta$ is related to the Landau--Zener
transition probability by $v=\sin^2{\theta/2}$.
Note that the expression (\ref{U-n}) is correct only to the leading
orders in the small parameters $w$ and $v$ and should be
treated as such.

Eq. (\ref{evolution}) must be solved for a stationary solution 
with fluctuating $\delta\varphi_n$. 
Since the phase fluctuations $\delta\varphi_n$ are assumed to
be uncorrelated, averaging over these fluctuations amounts
to averaging the evolution operator $Q_n$. 
After averaging $R_z(\delta\varphi_n)$
\begin{equation}
\overline{R_z(\delta\varphi_n)}=
\left(
\begin{array}{ccc}
e^{-u/2} & 0 & 0 \\
0 & e^{-u/2} & 0 \\
0 & 0 & 1 
\end{array}
\right)\ ,
\end{equation}
we arrive at the equation on the stationary solution 
$\bar{\bf a}$:
\begin{equation}
\label{evolution-averaged}
\bar{\bf a} = \bar{Q} \bar{\bf a} + w \hat{\bf z},
\quad
\bar{Q}=\overline{R_z(\delta\varphi_n)} R_z(\varphi) R_x(\theta) A_w\ .
\end{equation}

Solving (\ref{evolution-averaged}) for $\bar{\bf a}$, we 
find the population of the upper level $P_+$:
\begin{equation}
P_+(\varphi)={1{-}a_z \over 2} = {v(u+w) \over
{1\over 4} w (u+w)^2 + 2 v (u+w) + 4 w \sin^2 {\varphi\over 2}},
\label{P+}
\end{equation}
where we have kept only the leading terms in the small parameters
$w$, $u$, and $v$.

This equation describes Lorentzian peaks positioned at $\varphi=2\pi n$.
The peaks are sharp if
\begin{equation}
w \gg u v,
\label{strong-interference-condition}
\end{equation}
in which case the width of the peaks $\delta\varphi$ is given by
\begin{equation}
(\delta\varphi)^2={(u+w)^2\over4}+{2v(u+w)\over w}\ .
\label{peak-width}
\end{equation}

In the case of an arbitrary non-zero offset $x_{\rm off}$ superimposed onto
the periodic variation of the parameter (\ref{periodic-x}),
the two half-periods of the parameter sweep are no longer equivalent. 
The phase differences gained on odd and even half-periods
differ by the corresponding phase offset
$\varphi_{\rm off}=2t_0x_{\rm off}/\pi$:
$\varphi_n=\varphi+\delta\varphi_n\pm\varphi_{\rm off}$,
with the plus and minus sign for even/odd half-periods
respectively. As a consequence, the period of the master
equation (\ref{master-equation}) doubles, as it now includes two
half-periods of the parameter sweep. In the interference
pattern this produces secondary interference
peaks at $\varphi=\pi+2\pi n$. The relative intensities
of the two peaks depend on the offset $\varphi_{\rm off}$,
with the two intensities equal at $\varphi_{\rm off}=\pi/2+\pi n$,
and with one of the two peaks disappearing at 
$\varphi_{\rm off}=\pi n$. 

A tedious but straightforward calculation results in the following
extension of the formula (\ref{P+}) to the case of arbitrary
$\varphi_{\rm off}$:
%

%\end{multicols}
%\top
%
\begin{equation}
P_+(\varphi)=
{v(u+w)\left[1+\cos\varphi\,\cos\varphi_{\rm off}+{1\over 8}(u+w)^2\right]
\over D(\varphi, \varphi_{\rm off})}
\ , 
\label{P+off}
\end{equation}
where 
\begin{eqnarray}
\label{P+off-aux}
&&D(\varphi, \varphi_{\rm off}) = 
{1\over2}w(u+w)^2  + 2w \sin^2{\varphi}\\
&&+2v(u+w)\left[1+\cos\varphi\,\cos\varphi_{\rm off}+{1\over 8}(u+w)^2\right] 
\nonumber\ .
\end{eqnarray}
%
%\bottom
%\begin{multicols}{2}
%

This expression is again valid only to the leading orders in the
small parameters $u$, $v$, and $w$ 
(and coincides with Eq.~(\ref{P+}) for $\varphi_{\rm off} = 0$
only in this limit). 
The terms ${1\over8}(u+w)^2$
in the numerator and in the denominator are relevant only
near the points $1+\cos\varphi\, \cos\varphi_{\rm off} = 0$;
away from these points, the terms  ${1\over8}(u+w)^2$ are beyond
the precision of Eq.~(\ref{P+off}) and should be disregarded.
Several examples of interference curves at different
values of $\varphi_{\rm off}$ are plotted in Fig.~2. 
Provided the condition of strong interference 
(\ref{strong-interference-condition}) is satisfied, the height of 
secondary peaks in (\ref{P+off}) become equal to the background at
\begin{equation}
\varphi_{\rm off} \approx {u+w \over 2}.
\label{peaks-appear}
\end{equation}
Note that the secondary peak is much narrower than the 
primary one as long as its height is small: the width of 
small peaks is determined solely by the strength of decoherence
processes $u+w$ (only the first term in Eq.~(\ref{peak-width})),  
whereas the width of the high primary peak involves 
the Landau-Zener amplitude $v$.

So far our discussion assumed the reservoir temperature $T_{\rm res}$
much lower than the gap $\Delta$. Taking into account a finite
reservoir temperature, the inelastic processes in (\ref{inelastic})
must include not only transition from the upper level to the lower
one, but also the reverse transitions from the lower level to the
upper one (absorbing energy from the reservoir). The single transition
probability $w$ should then be replaced by the two probabilities
$w_1$ and $w_2$. Eq.~(\ref{inelastic}) is replaced by
\begin{equation}
\rho \mapsto V_w\rho V_w + w_1 \rho^{(0)}_1 \rho_{22}
+ w_2 \rho^{(0)}_2 \rho_{11} \ , 
\label{inelastic-3}
\end{equation}
where
\begin{eqnarray}
V_w &=& \left(
\begin{array}{cc}
    \sqrt{1-w_2}    & 0 \\
    0    & \sqrt{1-w_1}
\end{array}
\right)\ , 
\nonumber\\
\rho^{(0)}_1 = 
\left( 
\begin{array}{cc}
1 & 0  \\
0 & 0
\end{array}
\right)\ ,&&
\quad
\rho^{(0)}_2 = 
\left( 
\begin{array}{cc}
0 & 0  \\
0 & 1
\end{array}
\right) .
\label{inelastic-4}
\end{eqnarray}
We give the expressions
for the transition probabilities $w_1$ and $w_2$ in terms 
of the environment spectral function in the appendix. %~\ref{appendix-A}. 
Repeating the same derivation as before, we arrive to the equation
\begin{equation}
{\bf a}_{n+1}= Q_n {\bf a}_{n} + (w_1-w_2) \hat{\bf z}
\label{evolution-2}
\end{equation}
replacing Eq.~(\ref{evolution}), with $Q_n$ given by the same
expressions (\ref{U-n}) and (\ref{R-A}), except that now
\begin{equation}
A_w = \left(
\begin{array}{ccc}
\scriptstyle
\sqrt{(1-w_1)(1-w_2)} & 0 & 0 \\
0 & 
\scriptstyle
\sqrt{(1-w_1)(1-w_2)} & 0 \\
0 & 0 & 
\scriptstyle{
1-w_1-w_2} 
\end{array}
\right)\ . 
\end{equation}
For small $w_1$ and $w_2$, this expression for $A_w$ is equivalent
to introducing the effective decay probability $w=w_1+w_2$. Then
the solutions may be obtained from our previous low-temperature
results by a simple rescaling 
$\bar{\bf a}\mapsto \bar{\bf a} (w_1-w_2)/(w_1+w_2)$. In terms
of the average occupation number of the upper level $P_+(\varphi)$,
this translates to
\begin{equation}
P_+(\varphi)=P_+(\varphi,w=w_1+w_2) {w_1-w_2 \over w_1+w_2}
+{w_2 \over w_1+w_2},
\end{equation}
where $P_+(\varphi,w)$ in the right-hand side is given by 
Eqs.~(\ref{P+}) or (\ref{P+off}). In other words, at finite reservoir
temperatures, the interference pattern is simply rescaled by
the factor $(w_1-w_2)/(w_1+w_2)$. At low reservoir temperatures,
$T_{\rm res}\ll\Delta$, the ratio $w_2/w_1$ becomes exponentially small:
$w_2/w_1 \sim \exp(-2\Delta/T_{\rm res})$. At high reservoir
temperatures $T_{\rm res}\gg \Delta$, the probabilities $w_1$ and
$w_2$ are close to each other, 
$(w_1-w_2)/(w_1+w_2) \sim \Delta/T_{\rm res}$, which accordingly
decreases the amplitude (but not the sharpness) of the interference
pattern $P_+(\varphi)$.

Experimentally, it may be possible to measure the energy absorption
which is proportional to the population of the upper level $P_+$.
By observing the appearance of the secondary peaks at varying 
$\varphi_{\rm off}$, it should be possible from (\ref{peaks-appear})
to determine the combined decoherence rate $u+w$. This is precisely
the quantity which defines the quality of the qubit. 
The condition (\ref{strong-interference-condition}) should be fulfilled in order
resolve well interference picture. Estimating $w \sim \Gamma_{\rm relax} t_0\Delta/x_0$ and
$u \sim \Gamma_\phi t_0$,  (examples of estimates for longitudinal and transverse
relaxation rates   
 $\Gamma_{\rm relax}$ and $\Gamma_\phi$ for superconductive qubits can  be found
in \cite{Makhlin1,Makhlin2} ), and using (\ref{LZ-transition-probability}),
 one finds the condition
\begin{equation}
\frac{\Gamma_{\rm relax}}{\Delta} \gg \frac{\Gamma_\phi x_0}{\Delta^2}
\exp(-\frac{\Delta^2 t_0}{2x_0})
\label{cond2}
\end{equation}
The condition (\ref{cond2}) should be fulfilled together with 
inequalities $u, w \ll 1$. All these conditions together are compatible for 
low enough dephasing rate $\Gamma_\phi$; taking for the sake of estimate
$\Gamma_\phi/\Delta = 10^{-3}$ and $\Gamma_{\rm relax} \leq \Gamma_\phi $ (cf. Ref.
\cite{Makhlin2} ), we find broad interval of allowed $x_0$ and $t_0$.
Experimentally, values of $\Gamma_\phi/\Delta \sim 10^{-2}$ were measured in the first
 superconductive qubits~\cite{nakamura,mooij}, whereas much smaller normalized 
dephasing rate of order $10^{-4}$ was achieved for a  non-quasiclassical device
studied in Ref.~\cite{vion}. 

 It may be  useful to perform measurements at different values
of the amplitude $x_0$ and period $t_0$ of the parameter sweep.
Both Landau--Zener transition probability $v$ and the inelastic decay
probability $w$ depend only on the velocity at the transition point.
If $x_0$ and $t_0$ are changed simultaneously so that $x_0/t_0$ is kept
constant, $v$ and $w$ should also remain constant. At the same time,
for short-range correlations of $\xi(t)$, the dephasing $u$
scales linearly with $t_0$: $u=t_0 \Gamma_\phi$.   Under these assumptions,
from measurements at different $x_0$ and $t_0$ it may be possible to
determine the dephasing rate $\Gamma_\phi$, and the transition
probabilities $v$, and $w$. 

We are grateful to J.E.Mooij for a very useful discussion.
This research was supported by 
NSF grant PHY99-07949, NWO-Russia collaboration grant,
Swiss NF, RFBR grant \# 01-02-17759, by Russian ministry of science,
 and by the program "Quantum Macrophysics" of Russian Aca\-demy of Science.

\begin{figure}
\includegraphics[scale=0.4]{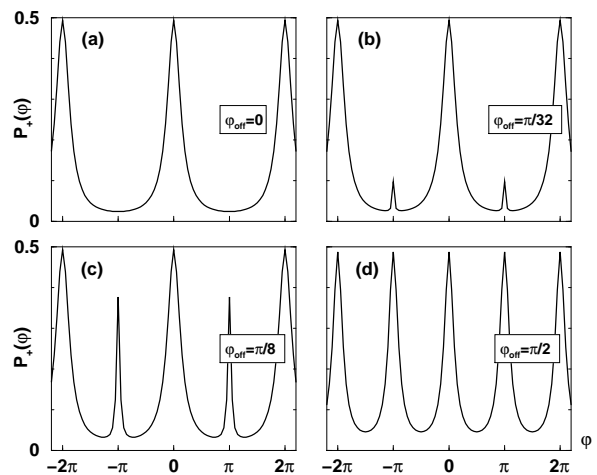}
\caption[]{
Interference pattern $P_+(\varphi)$ as given by Eqs.~(\ref{P+off})
and~(\ref{P+off-aux})
for different values of the offset $\varphi_{\rm off}$.
The non-adiabatic excitation probability  $v = 10^{-3}$, 
the decay probability $w = 10^{-3}$, the dephasing factor 
$u = 0.1$. The four curves {\bf (a)}--{\bf (d)} correspond to
$\varphi_{\rm off}=0$, $\pi/32$, $\pi/8$, and $\pi/2$ respectively.
}
\label{fig2}
\end{figure}

%\vspace{-1.0cm}

\appendix
\section*{Appendix: The microscopic derivation of the equation for the density matrix}
\label{appendix-A}

To establish connection between the microscopic
Hamiltonian (\ref{hamiltonian})
and the phenomenological Eqs.~(\ref{inelastic}) and (\ref{inelastic-3}),  
we  compute the density matrix directly. 
%, within perturbation
%theory.  
We consider the evolution of the qubit during one sweep, treating
the coupling to the environment perturbatively, find correction
to the density matrix $\hat{\rho}$, and compare it with the expansion 
of Eq.~(\ref{inelastic-3}) in small~$w_1$ and~$w_2$.  
%treating the coupling with 
%the environment as a perturbation. 
First, we rewrite the 
Hamiltonian~(\ref{hamiltonian}) in the basis of 
adiabatic states:
\begin{equation}
\hat{H} = \hat{H}_0(t) + \hat{V}(t) \ , 
\end{equation}
where the unperturbed Hamiltonian $\hat{H}_0(t)$ is diagonal, 
\begin{equation}
\label{zero-hamiltonian}
\hat{H}_0  = \hat{\sigma}_z \epsilon_t\ ; \qquad \epsilon_t = \sqrt{x^2(t) +
\Delta^2}\ , 
\end{equation}
and the perturbation $\hat{V}(t)$ is given by 
\begin{equation}
\label{perturbation}
\hat{V}(t) = \hat{\xi} (t) ( \cos\theta_t \hat{\sigma}_z 
+ \sin\theta_t \hat{\sigma}_x)\ ; 
\quad \theta_t = \tan^{-1}\frac{\Delta}{x(t)}
\ . 
\end{equation}
We neglect the probability of the LZ transition, and consider 
only the transitions due to the coupling to the environment 
$\hat{V}(t)$. In doing so, we assume that the
characteristic energies involved are or order of $\epsilon_t \sim \Delta$, 
and the main $t$--dependence of the perturbation $\hat{V}(t)$ 
is due to fast fluctuation of $\hat{\xi}(t)$. 
Since $\theta_t$ changes essentially only on the large time scale 
$\Delta t_0 / x_0$, we will treat it as a slow variable. 

To compute the evolution of the density matrix $\hat{\rho}$
under the Hamiltonian~(\ref{hamiltonian}), one may use Liouville equation
in Heisenberg representation
\begin{equation}
\dot{\hat{\rho}}(t) = i \left[\hat{\xi}(t)\hat{U}(t),  
                              \hat{\rho}(t)\right] \ , 
\end{equation}
where 
\begin{equation}
%\hat{\tilde{V}} (t) 
%= \hat{\xi}(t) \hat{U}(t)\ ; \qquad 
\hat{U}(t) =  \sin\theta_t 
e^{\frac{i}{2} \sigma_z \chi (t)} \sigma_x e^{-\frac{i}{2}\sigma_z\chi (t)} 
\ ; \quad  
\chi(t) = 2 \int\limits_{-\infty}^{t} \epsilon_t\, dt \ ,
\end{equation}
%In zero approximation, the evolution is simply
%\begin{equation}
%\hat{\rho}_{0}(t) \approx 
%\left(
%\begin{array}{ll}
%\rho_{11}  & \rho_{12} e^{i \chi (t)} \\
%\rho_{21} e^{-i\chi(t)} & \rho_{22}
%\end{array}
%\right) \ ;\qquad 
%with adiabatic phase 
%and
where $\chi(t)$ is the phase difference between the two adiabatic states. 
%$\chi(t)$ is
%computed within adiabatic approximation.
%\begin{equation}
%\chi(t) = 2 \int\limits_{-\infty}^{t} \epsilon_t dt \ .
%\end{equation}
%%\exp\left(i \int\limits^{t} \hat{H}_0(t) dt\right)
%%\hat{\rho} \exp\left(-i \int\limits^{t}\hat{H}_0 (t) dt \right)
%\end{equation}
The perturbation theory with respect to $V(t)$ gives, to the second order, 
\begin{eqnarray}
\label{perturbative-solution}
\hat{\rho} (t) &=& \hat{\rho}_0 (t) 
+ i \int\limits_{-\infty}^{t} dt_1\, 
    \left[\hat{\xi}(t_1)\hat{U}(t_1), \hat{\rho}_0 (t)\right]
\nonumber \\
&-& \int\limits_{-\infty}^{t}\!dt_1\!\int\limits_{-\infty}^{t_1}\!dt_2\,
\left[\hat{\xi}(t_1)\hat{U}(t_1), 
      \left[\hat{\xi}(t_2)\hat{U}(t_2), \hat{\rho}_0 \right]\right] 
\ , 
\end{eqnarray}
where $\hat{\rho}_0$ is the (time-independent) 
density matrix in zero approximation. 
Now, we average Eq.~(\ref{perturbative-solution}) over fluctuations of 
$\hat{\xi}(t)$. The first order term vanishes, and one has for 
$\delta\hat{\rho}(t) = \hat{\rho}(t) - \hat{\rho}_0$
\begin{eqnarray}
\label{micro-averaged}
\delta\hat{\rho}(t) = 
&-& \int\limits_{-\infty}^{t}\! dt_1 \int\limits_{-\infty}^{t_1}\! dt_2
\, \bigl[\hat{U}(t_1), 
          \hat{U}(t_2)\, Q(t_1 - t_2) \hat{\rho}_0
          \\
          &-& \hat{\rho}_{0} \hat{U}(t_2)\, Q(t_2 - t_1)\bigr]
          \ . \nonumber 
\end{eqnarray}
Here 
$
%\begin{equation}
Q(t) = \langle \hat{\xi}(t) \hat{\xi}(0) \rangle  
%\end{equation}
$ is the correlation function of the environment. 
Note that since $\hat{\xi}(t)$ is quantum variable, $Q(t) \neq Q(-t)$. 
Rewriting the commutator in Eq.~(\ref{micro-averaged}), one finds
\begin{eqnarray}
\label{pert-correction}
\delta\rho_{11}(t) &=& %- \delta\rho_{22}(t) = 
-  2\mathop{\rm Re}\nolimits
   \int\limits_{-\infty}^{t} dt_1 \int\limits_{0}^{\infty} d\tau\, Q(\tau)
   \, \sin\theta_t \sin\theta_{t - \tau} \times \\ 
    &\times& %\sin\theta(t - \tau) 
       \left(\rho_{11}e^{i \chi(t_1) - i \chi(t_1 - \tau)} 
           - \rho_{22} e^{i\chi(t_1 - \tau) - i \chi(t_1)} 
    \right) \ , \nonumber
\\
\delta\rho_{12}(t) &=& %\delta\rho^\ast_{21}(t) = 
 - \!\int\limits_{-\infty}^{t}\!dt_1\!\int\limits_{0}^{\infty}\!d\tau
  [Q(\tau) + Q(-\tau)] \sin\theta_{t} \sin\theta_{t-\tau}\\
  &\times& %\sin\theta_{t-\tau} 
          \left(\rho_{12}e^{i \chi(t_1 - \tau) - i \chi(t_1)} 
             - \rho_{21} e^{i\chi(t_1) + i \chi(t_1 - \tau)}\right) \ ,  
             \nonumber
%            \\
%  \delta\rho_{22} (t) &=& -\delta\rho_{11} (t) \ ; \qquad
%  \delta\rho_{21} (t) = \delta\rho^\ast_{12}(t) \ . 
\end{eqnarray}
and also $\delta\rho_{22} (t) = -\delta\rho_{11} (t) $, 
$\delta\rho_{21} (t) = \delta\rho^\ast_{12}(t) $.
The dominant contribution to the integral over $\tau$ in 
Eq.~(\ref{pert-correction}) comes from the region $\tau \sim
\epsilon^{-1}_t$, and one can use an approximation 
$\chi (t_1) - \chi (t_1 - \tau) \approx 2 \tau \epsilon_{t_1}$. 
For the oscillator bath 
\begin{equation}
\hat{H}_{\rm env} = \sum\limits_{i} \omega_i a_i^{+} a_i 
\ , \quad \hat{\xi}(t) = 
 \sum\limits_{i} \gamma_i (a_i e^{i \omega_i t} + a_i^{+} e^{-i\omega_i t})
\end{equation}
integrals in Eq.~(\ref{pert-correction}) can be expressed in terms 
of the environment spectral function~\cite{Caldeira-Leggett}
\begin{equation}
J(\Omega) = \sum\limits_i \frac{\gamma_i^2}{\omega_i} \, 
\delta (\Omega - \omega_i)
\end{equation}
as
\begin{eqnarray}
\int\limits_{0}^{\infty} Q(\tau) e^{i \omega \tau} d\tau %= %\\
%\nonumber 
&=& 2i \int\limits_{0}^{\infty} \Omega J(\Omega) d\Omega 
\left[
   \frac{1 + n (\Omega)}{\omega + \Omega - i 0} \right. \nonumber\\ 
 &-& \left.\frac{n(\Omega)}{\omega - \Omega - i0}
\right]
\ , 
\end{eqnarray}
where $n(\Omega) = (\exp(\Omega / T_{\rm res}) - 1)^{-1}$ 
is Bose-Einstein distribution function, and $T_{\rm res}$ is the
reservoir temperature. 

After a straightforward calculation, one arrives to the 
correction to the density matrix at the end of one sweep:
\begin{eqnarray}
\label{perturbative-final}
\delta \rho_{11}(\infty) &=& - \rho_{11} w_2 + \rho_{22} w_1 \\
\delta \rho_{12}(\infty) &=& -  \frac{1}{2}\, (w_1 + w_2) \rho_{12} 
                          + i \Phi \rho_{12}
\ . 
\end{eqnarray}
Here 
\begin{eqnarray}
w_1 &=& 4\pi \int\limits_{-\infty}^{\infty} \left[1 + n(2\epsilon_t)\right]
J (\epsilon_t) \sin^2\theta_t \, \epsilon_t\, dt \\
w_2 &=& 4\pi \int\limits_{-\infty}^{\infty} n(2\epsilon_t)
J (\epsilon_t) \sin^2\theta_t \, \epsilon_t \,dt \
\end{eqnarray}
are the transition probabilities, and 
\begin{equation}
\Phi = \int\limits_{-\infty}^{\infty} \epsilon_t \, dt \sin^2\theta_t
\int\limits_{0}^{\infty} \frac{[2 n (\Omega) + 1] J (\Omega) \Omega d\Omega} {
                               \epsilon_t^2 - \Omega^2}
\end{equation}
is the additional phase picked up during the sweep. This phase 
shift is due to the renormalization of the gap $\Delta$ due to 
the interaction between the qubit and the environment. 

Comparison of Eqs.~(\ref{perturbative-final}) and~(\ref{inelastic-3})
shows that the phenomenological Eq.~(\ref{inelastic-3}) is correct 
in the perturbative limit. Also, since $w_1$ and $w_2$ 
contain~$1 + n(2\epsilon_t)$ and $n(2\epsilon_t)$ respectively, 
in the low-temperature limit ($T_{\rm res} \ll \Delta$) the decay rate $w_1$
is finite, while the excitation rate $w_2$ is thermally assisted: 
$w_2 \sim \exp(-2\Delta / T_{\rm res})$. 
%one can see that the qualitative picture presented in our paper is correct. 

%\end{equation}

%\end{multicols}

\end{document}